 \documentclass{emulateapj}

\slugcomment{Not to appear in Nonlearned J., 45.}

\shorttitle{Gamma-ray Polarization in GRB~100826A}
\shortauthors{Yonetoku et al.}

\begin{document}


\title{Detection of Gamma-Ray Polarization in Prompt Emission of GRB~100826A}


\author{
Daisuke Yonetoku\altaffilmark{1},
Toshio Murakami\altaffilmark{1}, 
Shuichi Gunji\altaffilmark{2}, 
Tatehiro Mihara\altaffilmark{3},
Kenji Toma\altaffilmark{4}
Tomonori Sakashita\altaffilmark{1}, 
Yoshiyuki Morihara\altaffilmark{1}, 
Takuya Takahashi\altaffilmark{1}, 
Noriyuki Toukairin\altaffilmark{2}, 
Hirofumi Fujimoto\altaffilmark{1}, 
Yoshiki Kodama\altaffilmark{1}, 
Shin Kubo\altaffilmark{5},
and
IKAROS Demonstration Team\altaffilmark{6}
}
\email{yonetoku@astro.s.kanazawa-u.ac.jp}


\altaffiltext{1}{College of Science and Engineering, 
School of Mathematics and Physics,
Kanazawa University, Kakuma, Kanazawa, Ishikawa 920-1192, Japan}
\altaffiltext{2}{Department of Physics, Faculty of Science, 
Yamagata University, 1-4-12, Koshirakawa, Yamagata, Yamagata 990-8560, Japan}
\altaffiltext{3}{Cosmic Radiation Laboratory, RIKEN, 2-1, Hirosawa, 
Wako City, Saitama 351-0198, Japan}
\altaffiltext{4}{Department of Earth and Space Science, 
Osaka University, Toyonaka 560-0043, Japan}
\altaffiltext{5}{Clear Pulse Co. Ltd., 6-25-17, Chuo, Ohta-ku, 
Tokyo 143-0024, Japan}
\altaffiltext{6}{Institute of Space and Astronautical Science (ISAS), 
Japan Aerospace Exploration Agency (JAXA), 
3-1-1, Yoshinodai, Sagamihara, Kanagawa 229-8510, Japan}

\begin{abstract}
We report the polarization measurement in prompt $\gamma$-ray emission 
of GRB~100826A with the Gamma-Ray Burst Polarimeter (GAP) aboard 
the small solar power sail demonstrator IKAROS.
We detected the firm change of polarization angle (PA) during the prompt 
emission with 99.9~\% ($3.5~\sigma$) confidence level, 
and the average polarization degree ($\Pi$) of $27 \pm 11$~\% 
with 99.4~\% ($2.9~\sigma$) confidence level. Here the quoted errors 
are given at 1~$\sigma$ confidence level for two parameters of interest. 
The systematic errors have been 
carefully included in this analysis, unlike any previous reports.
Such a high $\Pi$ can be obtained in several emission models of 
gamma-ray bursts (GRBs), including synchrotron and photospheric models.
However, it is difficult to explain the observed 
significant change of PA within the framework of axisymmetric jet 
as considered in many theoretical works.
The non-axisymmetric (e.g., patchy) structures of the magnetic fields 
and/or brightness inside the relativistic jet are therefore required 
within the observable angular scale of $\sim \Gamma^{-1}$.
Our observation strongly indicates that the polarization measurement is 
a powerful tool to constrain the GRB production mechanism, and more 
theoretical works are needed to discuss the data in more details.
\end{abstract}



\section{Introduction}
Gamma-ray bursts (GRBs) are the most energetic explosions in
the universe, and the isotropic luminosity
reaches $10^{54}~{\rm erg~s^{-1}}$ for the brightest bursts.
Since the discovery of X-ray afterglow of GRBs by
BeppoSAX \citep{costa1997} and the identification of optical 
counterparts at cosmological distance, many observational facts 
and theoretical works have led us to understand the nature of 
GRBs. However, a crucial issue still remaining to answer is 
how to release such a huge energy as $\gamma$-ray photons. 
In spite of extensive discussions with the spectral and 
lightcurve information being collected, there are still 
several possible models of GRBs \citep{meszaros06}. 
The polarimetric observations provide us completely different 
information. A firm detection of linear polarization of 
$\gamma$-ray photons will further constrain the emission models
\citep{lazzati2006, toma2009}.

The polarization measurement of GRBs has been performed in 
the late optical afterglows at first
\citep{covino1999, wijers1999, rol2000, gorosabel2004, covino2004}. 
All of these results show that
the polarization degree ($\Pi$) of a few~\% level.
\citet{mundel2007} set upper limit of $\Pi < 8$~\% for 
the early optical afterglow.
Detailed observations are performed for the optical 
afterglow of GRB~030329, and time variations of 
the polarization degree and angle (PA)
were reported \citep{greiner2003}. 
Recently, a polarization degree of $\Pi = 10 \pm 1$~\% was reported 
for the early optical afterglow of GRB~090102,
at hundreds of seconds since the burst trigger \citep{steele2009}. 

On the other hand, there are controversies on polarization detections 
in the prompt $\gamma$-ray emission of GRBs. The first report was 
the detection of a high linear polarization of $\Pi = 80 \pm 20$~\% 
at a confidence level of $> 5.7~\sigma$ by RHESSI 
from GRB~021206 \citep{coburn2003}, although independent 
groups could not confirm any polarization signals by reanalyzing the
same data \citep{rutledge2004, wigger2004}. The second reports showed 
also high degree of polarization for GRB~041219A with
$\Pi = 98 \pm 33$~\% \citep{kalemci2007} and
$\Pi = 63^{+31}_{-30}$~\% \citep{mcglynn2007} for the brightest 
pulse in the 100--350~keV energy band of INTEGRAL-SPI at $2~\sigma$ 
confidence level for 2 parameters of interest ($\Pi$ and PA).
\citet{gotz2009} also reported more significant detection for 
time resolved analyses\footnote{See Table~1 and Figure~3 of 
\citet{gotz2009}. One should note that their definition of 
confidence level is given for one parameter of interest.}, 
and suggested possible variable polarization 
properties for the same GRB~041219A 
in the 200--800~keV energy band observed with INTEGRAL-IBIS.
However, \citet{gotz2009} set a strict 
upper limit of $\Pi < 4$~\% for the same brightest pulse showing 
high $\Pi$ in the INTEGRAL-SPI data \citep{kalemci2007, mcglynn2007}. 
(\citet{gotz2009} explained that the null net polarization degree
results from the superposition of signals with different 
polarization angles. However, this interpretation cannot explain 
the positive detection by INTEGRAL-SPI simultaneously.) 
These inconsistent results between SPI and IBIS led to confusion again, 
and the existence of $\gamma$-ray polarization is still in debate. 
As the authors themselves pointed out in their reports, 
these results are with low statistics of $\sim 2~\sigma$ level, 
and may be strongly affected by the instrumental systematics 
uncertainties \citep{kalemci2007, mcglynn2007, gotz2009}. 
These controversies and conflicts indicate difficulties in detecting 
the $\gamma$-ray polarization.

In this Letter, we report the detection of $\gamma$-ray polarization 
and also the change of PA for extremely bright 
GRB~100826A using the newly developed GRB polarimeter.
Our polarimeter is completely designed for the polarization 
measurement of prompt GRBs, and well calibrated during developing 
phase before launch. Especially, using a detector of proto-flight 
model, we experimentally and numerically understand the response 
for polarized $\gamma$-ray with low systematic uncertainties.
In the following sections, we show the observation (section~2), 
data analysis (section~3) and discussion for the emission mechanism 
of the prompt GRBs (section~4).
 
\section{Observations}
IKAROS \citep{kawaguchi2008, mori2010} 
(Interplanetary Kite-craft Accelerated by Radiation Of the Sun) 
is a small solar-power-sail demonstrator and was successfully
launched on 21 May 2010. IKAROS has a large polyimide membrane of 
20~m in diameter, and this translates the solar radiation pressure 
to the thrust of the spacecraft. Since the deployment of the sail 
on 9 June 2010, IKAROS started the solar-sailing toward Venus. 

The Gamma-Ray Burst Polarimeter (GAP) 
\citep{yonetoku2011, murakami2010, yonetoku2006} 
aboard IKAROS is fully designed to measure a degree of linear 
polarization in the prompt emission of GRBs 
in the energy range of 70--300 keV. The detection principle is
an anisotropy of Compton scattered photons.
If the incident $\gamma$-rays are linearly polarized, the distribution 
of scattered photons is due to Klein-Nishina effect 
which approximately shows $\sin^{2} \phi$ curves, 
where $\phi$ is the scattering angle. The GAP consists of 
a central plastic scatterer of 17~cm in diameter and 6~cm in thickness, 
and the surrounding 12 CsI(Tl) scintillators with 5~mm in thickness. 
The coincidence events within a gate time of $5~{\rm \mu s}$ 
between the signal from any CsI and that from the plastic are accepted 
for the polarization analysis. The GAP's high axial symmetry in shape 
and the high gain uniformity are the keys for reliable
measurement of polarizations, and to avoid a fake modulation due to
background $\gamma$-rays. 
There are no external parts of spacecraft inside the GAP field of view. 
Moreover the detector cases (chassis), except for the detector top,
are covered by thin lead sheets with 0.5~mm thickness.
Therefore the effect of background $\gamma$-rays scattered by 
the spacecraft body are negligible.

\begin{figure}
\includegraphics[angle=270,scale=0.50]{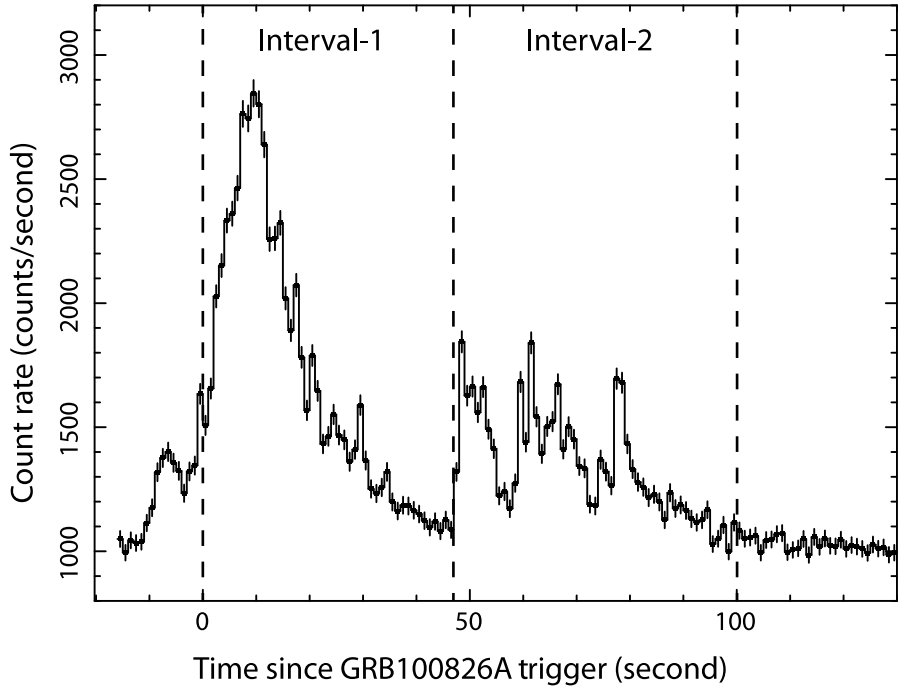}
\caption{Lightcurve of the prompt $\gamma$-ray emission of GRB100826A
detected by the GAP. 
We divide the data into Interval-1 and -2 for the polarization analysis.
\label{fig1}}
\end{figure}

The GAP detected GRB~100826A on 26 August 2010 at 22:57:20.8 (UT) 
on the way to Venus at about 0.21~AU apart from the Earth. 
The lightcurve of the prompt emission is shown in Figure~\ref{fig1}. 
This burst was also detected by several satellites, and was 
localized by an interplanetary network (IPN) \citep{hurley2010}. 
Combining the GAP data with the published IPN information, 
the direction of this burst is derived as
$(\alpha, \delta) = (279.6 \pm 0.2, -22.3 \pm 0.5)$, 
which corresponds to 20.0~degree off-axis from the center of 
the GAP field of view. An energy fluence of
this burst is $(3.0 \pm 0.3) \times 10^{-4}~{\rm erg~cm^{-2}}$ 
in 20~keV--10~MeV band by KONUS \citep{golenetskii2010}, 
which is the top 1~\% of the brightest events listed in BATSE
catalog. 
The low- and high-energy photon indices are reported as
$\alpha_B = -1.31^{+0.06}_{-0.05}$ and $\beta_B = -2.1^{+0.1}_{-0.2}$,
respectively, and the $\nu F_{\nu}$ peak energy as 
$E_p = 606^{+134}_{-109}\;$keV \citep{golenetskii2010}.
An optical afterglow of this GRB was not
reported, so its redshift is unknown.

\section{Data Analysis}
We divided the entire data into two time intervals as labeled 
Interval-1 and -2 in a lightcurve of Figure~\ref{fig1}. 
The total numbers of $\gamma$-ray photons after subtracting background 
are 32,924 and 19,007 photons for each interval, respectively. 
The first part of this burst shows a large flare lasting 
47~seconds since the trigger, and a following 53~seconds consists 
of multiple spikes. Although the Interval-2 has several spikes, 
we combined all of them to keep photon statistics.

We used the background modulation curve obtained by 
the 36.7~hours integration just before and after the GRB 
trigger time. In this case, the averaged coincidence background 
rate is quite stable as $5.6~{\rm counts~s^{-1}CsI^{-1}}$, 
and its modulation curve can be described as constant with 
the standard deviation of 0.1~\% level. After subtracting 
the background modulation, the total numbers of the coincidence 
$\gamma$-rays (polarization signals) are 4,821 and 2,733 for 
Interval-1 and -2, respectively. 

\subsection{Model Functions}
When we perform the polarization analysis, we calculated 
the model function (or the detector response) for the polarized 
gamma-rays with Geant~4 Monte-Carlo simulator. 
A geometrical mass model of GAP was introduced in 
\citet{yonetoku2011}. We can set all the possible situations
with different off-axis angles, azimuthal phase angles, spectral
parameters, as well as polarization degrees and angles of incoming 
$\gamma$-rays.
In this analysis, the free parameters are the polarization 
degrees and angles, since the other parameters are constrained
as shown in section~2.

Simulating the interactions between the incoming gamma-rays 
and the geometrical mass model with Monte-Carlo method, 
we created the model functions of the modulation curves. 
Then we selected the coincidence events between the plastic and 
one CsI scintillator as the polarization signals. 
We simulated the model function with the step resolutions of 5~\% 
for polarization degrees and 5~degrees for phase angles. 
We will fit the observed modulation curves with the model 
function by a least squares method in the following subsections.

\subsection{Systematic Uncertainty}
We should treat not only the statistical error but also 
the systematic uncertainties correctly. Here, the systematic 
uncertainties mean any unexplainable difference between 
the observed modulation curve and the simulated one. 

Imperfect tunings of parameters in the ground and in-orbit 
calibrations for the incident radiation, especially from 
off-axis direction, cause more important systematic uncertainty.
First, we performed some ground based experiments 
with non-polarized radio isotopes $^{57}$Co (122~keV) and 
$^{241}$Am (59.5~keV). They were set at the distance of 1.0~m 
from the center of proto-flight model with several 
incident angles from zero to 50~degrees with the step resolutions 
of 5~degrees. We measured experimental modulation curves 
for each setup. After that, on the assumption of non-polarized 
$\gamma$-rays with the monochromatic energy of 122~keV and/or 59.5~keV, 
we calculated the numerical modulation curve with Geant~4 simulator.

Then, of course, the both modulation curves show some discrepancy 
which cannot be explained within the statistical uncertainty,
because of instrumental systematic uncertainties.
We calculated the standard deviation of two modulations 
and derived the systematic errors to be $\sigma_{sys} = 1.8$~\% 
of the total coincidence $\gamma$-rays for the case of 20.0~degree 
incident angle. Although this value is not negligible, 
the statistical error (about $\sigma_{stat} \sim 5$~\% leves) 
still dominates the systematic one in this case. 
We included the systematic uncertainty into the total errors as 
$\sigma^{2}_{total} = \sigma^{2}_{stat} + \sigma^{2}_{sys}$
for each bin of polarization data.

\subsection{Fitting the Polarization Data}
\begin{figure}
\includegraphics[angle=0,scale=0.60]{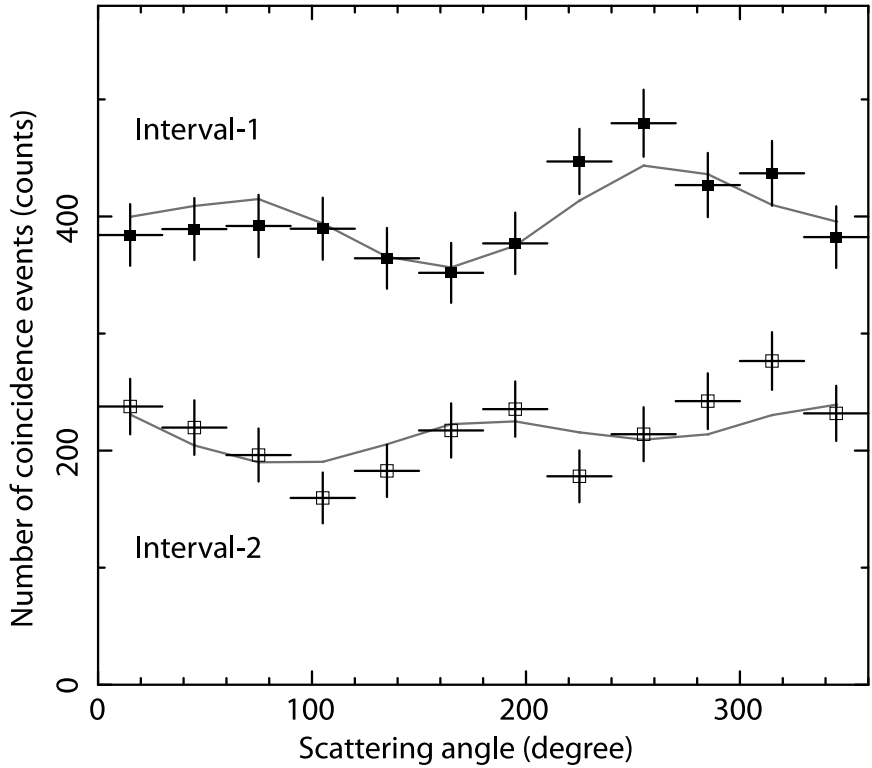}
\caption{Number of coincidence $\gamma$-ray photons 
(polarization signals) against 
the scattering angle of GRB~100826A measured by the GAP 
in 70--300~keV band. Black filled and open squares are 
the angular distributions of 
Compton scattered $\gamma$-rays of Interval-1 and -2, respectively. 
The gray solid lines are 
the best-fit models calculated with our Geant4 Monte-Carlo simulations. 
\label{fig2}}
\end{figure}

First of all, we performed the polarization analysis for 
the whole dataset of time interval of 0--100~seconds.
Then we obtained an acceptable result with the model of 
non-polarized modulation curve, and set only upper limit of 
$\Pi < 30$~\% ($2~\sigma$ confidence level). We consider the 
possibility that the polarization degree may be weak if 
the polarization angle changes during the entire burst duration,
as suggested by \citet{gotz2009}.
Therefore, we performed time resolved polarization analyses 
for datasets of Interval-1 and -2.
In Figure~\ref{fig2}, 
we show the observed source modulation curves as a function of 
scattering angle for both time intervals after subtraction of 
background. The error bars shown in Figure~\ref{fig2} include 
not only the statistical uncertainty but also the 1.8~\% of 
systematical one as estimated the above subsection.

At the first step, we investigated the polarization degrees, $\Pi$, 
and the polarization angle ($\phi_p$), measured from north with 
anticlockwise direction, separately for Interval-1 and -2. 
Then, the response of GAP for irradiation from 20.0~degree 
off-axis is modeled by the Monte-Carlo method with Geant~4 
simulator. The gray solid lines are the best-fit functions 
for Interval-1 and -2. 
The best values are $\Pi_1 = 25 \pm 15$~\% with 
$\phi_{p1} = 159 \pm 18$ degrees for Interval-1 and 
$\Pi_2 = 31 \pm 21$~\% with $\phi_{p2} = 75 \pm 20$ degrees
for Interval-2, respectively. Hereafter the quoted errors are 
$1~\sigma$ confidence for two parameters of interest. 
The significance of polarization detection 
is rather low of 95.4~\% and 89.0~\% for Interval-1 and -2, 
while the difference of polarization angles 
is significant with 99.9~\% ($3.5~\sigma$) level.

\begin{figure}
\includegraphics[angle=0,scale=1.0]{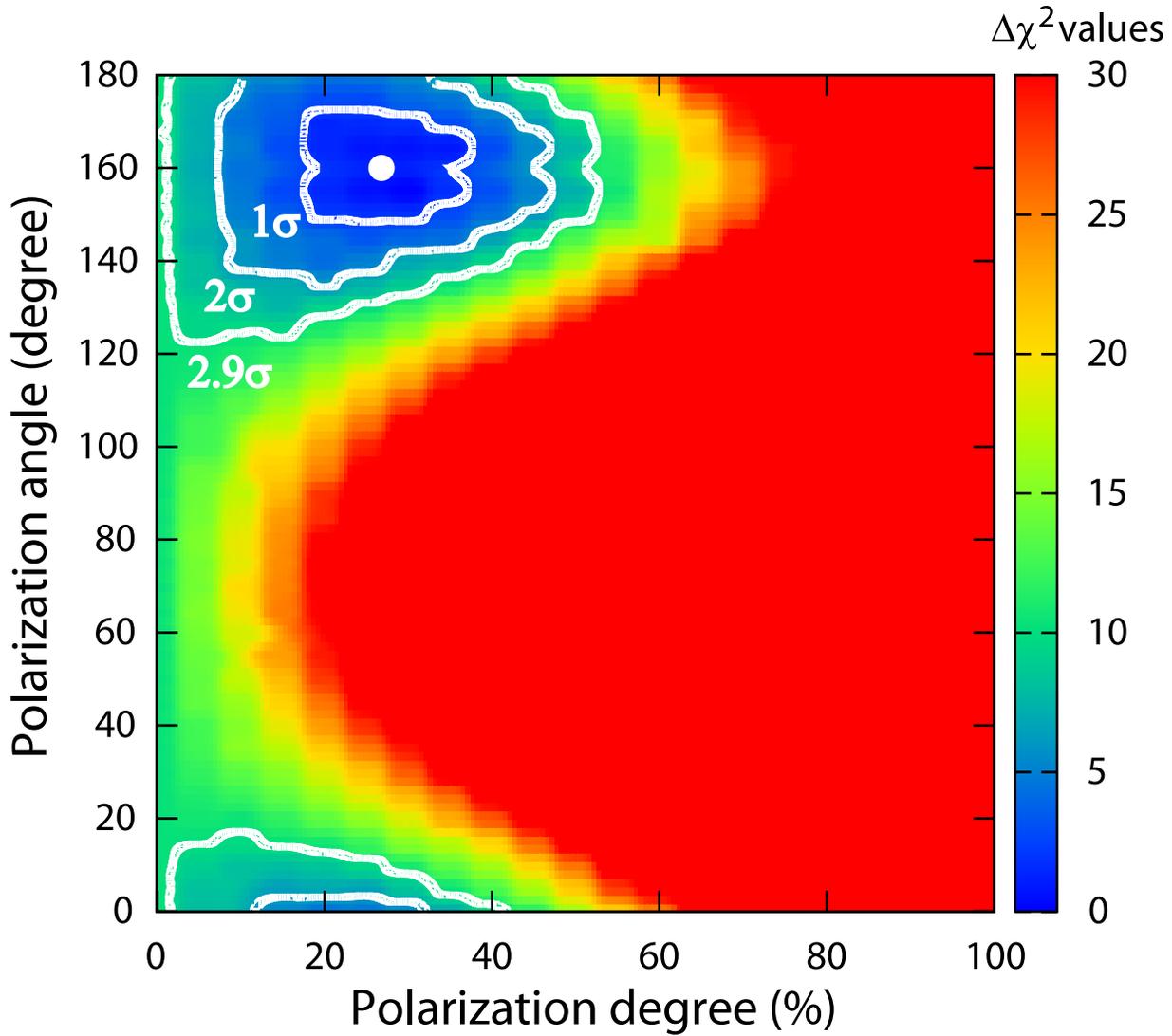}
\caption{A $\Delta \chi^{2}$ map of confidence contours 
in the $(\Pi, \phi_p)$ plane for GRB~100826A,
obtained by the combined fit of the Interval-1 and -2 data.
Here $\phi_{p}$ is the phase angle for Interval-1.
The white dot is the best-fit result, and we calculate 
$\Delta \chi^{2}$ values relative to this point. 
A color scale bar along the right 
side of the contour shows levels of $\Delta \chi^{2}$ values. 
The null hypothesis (zero polarization degree) can be ruled out 
with 99.4~\% ($2.9~\sigma$) confidence level.\label{fig3}}
\end{figure}

In the next step, we performed a combined fit for the two intervals, 
assuming that the polarization degree for Interval-2 is 
the same as that for Interval-1. This means that we treat $\Pi$ 
as one free parameter to improve the statistics with the reduction 
of model parameters. Here the two polarization angles were still 
free parameters for both intervals, because the change of angle 
is apparent. The best-fit polarization degree is 
$\Pi = 27 \pm 11$~\% with $\chi^{2} = 21.8$ for 19 degrees of freedom. 
We show a $\Delta \chi^{2}$ map in the $(\Pi, \phi_{p})$ plane in 
Figure~\ref{fig3}, where we represent $\phi_{p} = \phi_{p1}$. 
The significance of detection of polarization is
99.4~\% $(2.9~\sigma)$ confidence level. 

Gamma-ray detectors aboard RHESSI and INTEGRAL have complex 
geometry, the polarization analysis may be highly affected by 
the instrumental systematics as the authors mentioned
\citep{kalemci2007, mcglynn2007}. In contrast to them, 
GAP is developed for the purpose of GRB polarimetry with 
the high axial symmetry in shape. The systematic uncertainty 
is quantitatively estimated and found to be as small 
as 1.8~\% of the total polarization signals, as described above.
Of course, GAP is well calibrated for polarized 
gamma-rays before launch \citep{yonetoku2011}. 
Therefore, for the prompt emission of GRB,  we conclude that 
this is probably the most convincing detection of polarization 
degree and angles so far.

\section{Discussion} \label{discussion}
The emission mechanisms of GRBs which are being actively discussed 
with the spectral and lightcurve information collected so far 
includes synchrotron emission of electrons at 
the optically-thin regions of relativistic jets and 
quasi-thermal emission from the photospheres of jets 
\citep[for a review,][]{meszaros06}. Our reliable polarimetric 
observations may help further constrain the emission mechanisms 
as well as the structure of the emission sites and the magnetic 
fields. The main results of our polarimetric observation of 
GRB~100826A are: 
(1) The polarization degree averaged over the burst duration is
$\Pi = 27 \pm 11$~\% in the energy range of $70 - 300$~keV below 
$E_{p} \sim 600~{\rm keV}$; and 
(2) The PA significantly varies from Interval-1 to -2.

It has been theoretically shown that sizable net polarizations 
can be obtained both in the synchrotron and photospheric emission 
mechanisms. 
The synchrotron mechanism involves several different types of 
models with respect to magnetic field structure 
\citep[for reviews,][]{lazzati2006,toma2009};
(a) synchrotron in globally ordered toroidal field
\citep[SO model;][]{granot03,lyutikov03,zhang11}, 
(b) in random fields on plasma skin depth scales 
\citep[SR model;][]{granot03,nakar03}, and
(c) in random fields on hydrodynamic scales 
\citep[SH model;][]{inoue11,gruzinov99}. 
In the SR model, the magnetic field directions are assumed
to be random not isotropically but mainly in the plane perpendicular
to the local expansion direction. In this case
a high degree of polarization can be obtained only when the observer
sees the jet slightly off-axis, i.e.,
$\theta_v \sim \theta_j + \Gamma^{-1}$, 
where $\theta_v$ is the angle between the line of sight and 
the jet axis, and $\theta_j$ and $\Gamma$ are the opening half 
angle and the bulk Lorentz factor of the jet, respectively.
On the other hand, (d) in the photospheric emission model (Ph model),
we also expect the strongly polarized radiation
when the radiation energy is comparable to or smaller than 
the baryon kinetic energy at the photosphere \citep{beloborodov11}. 
Around the photosphere the photon distribution is highly anisotropic,
which can lead to high levels of polarization through the last 
electron scatterings 
\citep[see also][for the other Comptonization models]{eichler03,lazzati04}. 
We need the condition $\theta_v \sim \theta_j + \Gamma^{-1}$ also 
in this model to have a high net polarization.
The above four models can be consistent with our result (1). 

Our result (2) provides us with very important information, 
which excludes the axisymmetric jets both in the synchrotron and 
photospheric emission models. If the emitting surface is 
symmetric around the jet axis as assumed by most of authors in 
the SO, SR, and Ph models, the direction of the net polarization
in the high $\Pi$ case is determined to be either parallel or perpendicular 
to the direction from the jet axis towards the line of sight. 
Therefore we conclude that non-axisymmetric structure of the brightness 
and/or the magnetic fields on the observable angular scale 
$\sim \Gamma^{-1}$ are required to have the substantial change of PA
\citep[see also][]{lazzati09}.

A possible solution for the results (1) and (2)
is the patchy emission in the SO, SR, or Ph model, 
or the patchy magnetic field structure, i.e., the SH model. 
For instance, the SH model produces the net polarization as
$\Pi \sim 70\%/\sqrt{N}$, where $N$ is the number of independent
patches with coherent magnetic field in the observable region.
The polarization degree for $N \sim 2-50$ can be consistent with 
the individual results for Intervals-1 and -2 within 
the $1~\sigma$ confidence. 
The angle of the averaged polarization is statistically random,  
so that they can be consistent with our result (2). 
Similar arguments may be made in the patchy SO, SR, and Ph models.
In the patchy SO model, the emission from each patch 
can have different PA easily when $\theta_j \sim \Gamma^{-1}$.
In the patchy SR and Ph models, we no longer require the off-axis 
viewing of the jet, and the emission from each of off-axis viewed patches
has high $\Pi$ and different PA.

How can we further constrain the models with the analysis results of 
other bursts detected by GAP and the future polarimetry missions?
We note that the statistical analysis results of 
the SO, SR and Ph models in \citet{toma2009} are not applicable 
since they assume the axisymmetric jets. 
In the Ph model, the patch scales are constrained as 
$\theta_p \gtrsim \Gamma^{-1}$. Then high local polarizations 
are obtained from patches with 
$\theta_v \sim \theta_p + \Gamma^{-1} \gtrsim 2 \Gamma^{-1}$, 
although their brightness is much smaller than that from 
patches with $\theta_v \lesssim \Gamma^{-1}$ because of 
the relativistic beaming effect. Thus it is not expected that all 
the bright bursts have a high net polarization in 
the patchy Ph model. The statistical studies of the GAP bursts 
might provide us some implications for further constraining 
the emission mechanism.

This polarimetric observation for GRB~100826A has been performed 
in the energy range below $E_p$. In reality, the spectral indices 
$\alpha_B$ of bursts below $E_p$ are not fully explained 
in the framework of synchrotron or photospheric 
mechanism. In the synchrotron mechanism, the radiative cooling 
of emitting electrons should not be so rapid to produce 
the spectrum sufficiently hard below $E_p$, so that the emission 
region could be thick, unlike the thin shell emission 
assumed in many papers \citep[e.g.,][]{asano09}. 
The photospheric models include an idea that the emission 
below $E_p$ is superposition of many quasi-thermal components 
with different temperature and/or $\Gamma$ \citep[e.g.,][]{ryde10,toma11}. 
Clearly, more studies on the energy dependence of polarization
(as well as the non-axisymmetric structure of the jet)
in each model are needed to discuss the observational results 
quantitatively.

\acknowledgments

This work is supported by the Grant-in-Aid for 
Young Scientists (S) No.20674002 (DY), 
Young Scientists (A) No.18684007 (DY),
JSPS Research Fellowships for Young Scientists No.231446 (KT), and 
also supported by the Steering Committee for Space Science 
at ISAS/JAXA of Japan.
KT thanks K.~Ioka for useful conversations.

\clearpage

\end{document}